\documentclass[lettersize,journal]{IEEEtran}
\usepackage{amsmath,amsfonts}
\usepackage{algorithm} 
\usepackage{algorithmic}
\usepackage{array} 
\usepackage{textcomp}
\usepackage{stfloats}
\usepackage{color}
\usepackage{url}
\usepackage{verbatim}
\usepackage{graphicx}
\usepackage{verbatim}
\usepackage{bm}
\usepackage{amsthm,amssymb}
\usepackage{mathrsfs}
\usepackage{overpic}
\usepackage{cite}

\begin{document}

\title{Fluid Antenna System-Enabled Interference Alignment}

\author{Jialu Zheng, Quanzhong Li, and Qi Zhang, \emph{Member}, \emph{IEEE}

\thanks{\emph{(Corresponding author: Qi Zhang.)}

Jialu Zheng and Qi Zhang are with the School of Electronics and Information Technology, Sun Yat-sen University, Guangzhou 510006, China (e-mail: zhengjlu6@mail2.sysu.edu.cn; zhqi26@mail.sysu.edu.cn). Quanzhong Li is with the School of Computer Science and Engineering, Sun Yat-sen University, Guangzhou 510006, China (e-mail: liquanzh@mail.sysu.edu.cn).}

}

\markboth{}{Zheng \textit{et al.}: FAS-Enabled IA}

\maketitle

\begin{abstract}
Interference alignment is an efficient spectrum reuse scheme for multiuser wireless networks. In this letter,
assuming no channel state information at the transmitters, we investigate a fluid antenna system (FAS)-enabled interference alignment scheme over a multiuser interference channel. In the proposed system, the additional spatial degrees of freedom provided by the FAS are exploited to nullify the real components of the effective combined interference such that those are reserved for interference-free desired signal detection. The interference alignment conditions and corresponding outage probabilities are theoretically derived. Simulations verify that the theoretically derived outage probabilities agree well with the numerical results. Furthermore, the proposed system is able to achieve a higher average sum rate than conventional single-user transmission schemes over non-interfering channels.
\end{abstract}

\begin{IEEEkeywords}
Fluid antenna system (FAS), interference alignment, interference channels, outage probability.
\end{IEEEkeywords}

\section{Introduction}

For multiuser wireless networks with aggressive spectrum reuse, interference alignment coordinates transmit and/or receive signal processing such that multiple interference components occupy a lower-dimensional signal subspace, thereby reserving available dimensions for the desired signal \cite{Cadambe,Motahari}. Conventional interference alignment typically relies on channel-dependent transceiver beamforming or symbol extensions over time or frequency to construct the required signal spaces \cite{Cadambe,Motahari}. To mitigate the reliance on instantaneous channel state information at the transmitters, blind interference alignment exploits channel coherence patterns or structured channel variations to realize this alignment principle \cite{Jafar}. In particular, blind interference alignment schemes using receive pattern reconfigurable antennas to generate the required channel variations were proposed in \cite{Gou,Lu}. 

Whereas reconfigurable antennas create channel variations by switching among receive patterns, a fluid antenna system (FAS) exploits spatial channel variations by adjusting the antenna position \cite{KKWong21,TWu24}, thereby providing an alternative means of interference mitigation. In \cite{KKWong23}, for an FAS-aided multiple access, opportunistic user scheduling was proposed. In \cite{TWu26}, the performance of FAS and reconfigurable intelligent surface (RIS)-assisted 
non-orthogonal multiple access (NOMA) secure communication system was theoretically derived under realistic hardware constraints. In \cite{Peng}, Peng \emph{et al.} integrated fluid-antenna position optimization into blind interference alignment under imperfect channel state information and employed a learning-based method to maximize the robust sum rate. Besides FAS, adjusting antenna position for interference alignment and/or interference cancellation was also studied in \cite{Leshem,LZhu}.

In this letter, we investigate an FAS-enabled interference alignment system in a multiuser interference channel. In the proposed system, each transmitter is equipped with a single antenna and transmits signals to its intended receiver without any channel state information. Each receiver employs a one-dimensional linear FAS to nullify the real components of the effective combined interference, so that the corresponding real components of the received signals can be utilized for interference-free desired signal detection. We theoretically derive the aforementioned interference alignment conditions, based on which an exact outage probability expression is also theoretically derived.

\section{System Model}

Consider a $K$-user interference channel consisting of $K$ transmitters and $K$ receivers. Each transmitter is equipped with a single conventional antenna. Each receiver is equipped with an FAS that consists of $M$ fluid antennas. These antennas can move along a one-dimensional line segment of length $L$. Let $0 \leq x_{k,m}\leq L$ denote the position coordinate of the $m$th fluid antenna, $m\in\mathcal{M}=\{1,2,\cdots,M\}$, at the $k$th receiver, $k\in\mathcal{K}=\{1,2,\cdots,K\}$, satisfying the minimum spacing constraint \cite{KKWong21,TWu24}
\begin{equation}
\left|x_{k,m_1} - x_{k,m_2}\right| \geq \frac{\lambda}{2}
\end{equation}
for any $m_1\neq m_2$ and $m_1,m_2\in\mathcal{M}$, where $\lambda$ denotes the carrier wavelength.

The complex channel coefficient from the $i$th transmitter, $i\in\mathcal{K}$, to the $m$th fluid antenna, $m\in\mathcal{M}$, of the $k$th receiver, $k\in\mathcal{K}$,  is modeled as
\begin{equation}\label{q2}
h_{i,k}^{(m)} = g_{i,k}\exp\left(j2\pi x_{k,m} \cos\theta_{i,k}/\lambda\right)
\end{equation}
where $g_{i,k}$ denotes the complex propagation gain at the position with coordinate of 0, $\theta_{i,k}\in[0,2\pi)$ denotes the direction of arrival and $2\pi x_{k,m} \cos\theta_{i,k}/\lambda$ is the spatial phase shift due to the extra propagation delay at the position of $x_{k,m}$.

The received signal at the $m$th fluid antenna of the $k$th receiver is
\begin{equation}
y_k^{(m)} =w_k^{(m)}+\sum_{i=1}^K h_{i,k}^{(m)} s_i
\end{equation}
where $s_i \in \mathbb{R}$ denotes the transmitted symbol of the $i$th transmitter user intended for the $i$th receiver with $\mathbb{E}[|s_i|^2]=1$ and $w_k^{(m)} \sim \mathcal{CN}(0, \sigma^2)$ denotes the additive Gaussian noise.
 
The $k$th receiver combines the signals from its $M$ fluid antennas, which is expressed as $\sum_{m=1}^M y_k^{(m)}$.
For interference alignment \cite{Cadambe,Motahari}, we require that the real part of the combined output contains no interference from the $K-1$ undesired transmitters, which is used for desired signal detection. This condition is expressed as
\begin{equation}\label{q5}
\text{Re}\left(\sum_{m=1}^M h_{i,k}^{(m)} \right) = 0,\ \forall\ i \neq k.
\end{equation}
Substituting \eqref{q2} into \eqref{q5}, the interference alignment condition becomes
\begin{equation}
\sum_{m=1}^M \cos\left(\psi_{i,k} + \alpha_{i,k} x_{k,m} \right) = 0,\ \forall\ i \neq k
\end{equation}
where $\psi_{i,k}=\angle g_{i,k}$ denotes the phase of $g_{i,k}$ and $\alpha_{i,k} = \frac{2\pi}{\lambda} \cos\theta_{i,k}$ denotes the spatial frequency of the channel variation along the FAS. In the following section, we analyze how to satisfy the $K-1$ functions.

\section{Interference Alignment}

Without loss of generality, we consider the $K$th user as the desired user and solving the equations
\begin{equation}\label{bq1}
\sum_{m=1}^M \cos\left(\psi_{i} + \alpha_{i} x_m \right) = 0,\ \forall\ i\in\{1,\cdots,K-1\}
\end{equation}
where $\psi_{i}$ is the abbreviation for $\psi_{i,K}$, $\alpha_{i}$ is the abbreviation for $\alpha_{i,K}$, and $x_m$ is the abbreviation for $x_{K,m}$.

\subsection{The Condition of $M=1$}

When $M=1$, each receiver has only one fluid antenna at position $x_1$. The $K-1$ Equations in \eqref{bq1} are
\begin{equation}\label{bq2}
\cos\left(\psi_{i} + \alpha_{i} x_1 \right) = 0,\ \forall\ i\in\{1,\cdots,K-1\}.
\end{equation}
When $K=2$, the solution to \eqref{bq2} is
\begin{equation}
x_1= \frac{\pi/2 + n\pi - \psi_1}{\alpha_1}
\end{equation}
where $n\in\mathbb{Z}$ is an integer. When $K>2$, the $K-1$ Equations in \eqref{bq1} are overdetermined equation‌s, which generally have no solution. This is because both $\psi_{i}$ and $\alpha_{i}$ are independent random variables for $i\in\{1,\cdots,K-1\}$.

\subsection{The Condition of $M=2$ and $K=3$}

When $M=2$ and $K=3$, the interference alignment equations become sums of two cosine terms
\begin{align}
\cos(\psi_1 + \alpha_1 x_1) + \cos(\psi_1 + \alpha_1 x_2) &= 0,\\
\cos(\psi_2 + \alpha_2 x_1) + \cos(\psi_2 + \alpha_2 x_2) &= 0.
\end{align}
Using the sum-to-product formulas for cosine, we obtain
\begin{align}\label{bq11}
2\cos(\psi_1 + \alpha_1 u) \cos(\alpha_1 v) &= 0,\\
\label{bq12}
2\cos(\psi_2 + \alpha_2 u) \cos(\alpha_2 v) &= 0,
\end{align}
where
\begin{equation}\label{bq13}
u = \frac{x_2 + x_1}{2},\ v = \frac{x_2 - x_1}{2}.
\end{equation}
Without loss of generality, we assume $x_2>x_1$ and thus $u>v>0$. Each equation is satisfied if either of its two factors is zero. There exist the following 4 cases to satisfy the interference alignment equations \eqref{bq11} and \eqref{bq12}:
\begin{align}\label{bq14}
&\text{Case 1: }\cos(\psi_1 + \alpha_1 u) = 0 \text{ and } \cos(\alpha_2 v) = 0,\\
&\text{Case 2: }\cos(\alpha_1 v) = 0 \text{ and } \cos(\psi_2 + \alpha_2 u) = 0,\\
&\text{Case 3: }\cos(\psi_1 + \alpha_1 u) = 0 \text{ and } \cos(\psi_2 + \alpha_2 u) = 0,\\
&\text{Case 4: }\cos(\alpha_1 v) = 0\text{ and } \cos(\alpha_2 v) = 0.
\end{align}

Case 1 yields
\begin{align}\label{bq20}
u = \frac{n_1 \pi - \psi_1+\pi/2}{\alpha_1} \text{ and } 
v= \frac{n_2 \pi+\pi/2}{\alpha_2}
\end{align}
where $n_1,n_2\in\mathbb{Z}$ are integers. From \eqref{bq20}, $v$ is determined solely by the spatial frequency $\alpha_2$, while $u$ must also align with the phase $\psi_1$.

Case 2 yields
\begin{align}
v = \frac{n_1 \pi+\pi/2}{\alpha_1}  \text{ and }  u = \frac{n_2 \pi - \psi_2+\pi/2}{\alpha_2}
\end{align}
which is almost the same as Case 1 except that indices $1$ and $2$ swapped.

Case 3 requires
\begin{equation}\label{bq22}
v = \frac{n_1 \pi+\pi/2}{\alpha_1} = \frac{n_2 \pi+\pi/2}{\alpha_2}.
\end{equation}
It is noted that equation \eqref{bq22} involves only $v$ while the variable $u$ remains free. Since both $\alpha_1$ and $\alpha_2$ are independent random variables, equation
\eqref{bq22} generally has no solution. 

Case 4 requires
\begin{equation}\label{bq23}
u=\frac{n_1\pi-\psi_1+\pi/2}{\alpha_1} =\frac{n_2\pi-\psi_2+\pi/2}{\alpha_2}.
\end{equation}
Because both $\psi_{i}$ and $\alpha_{i}$ are independent random variables for $i\in\{1,2\}$, equation
\eqref{bq23} generally has no solution. 

From the analysis above, both Cases 3 and 4 have no solution. Cases 1 and 2 are almost the same except that indices $1$ and $2$ swapped.

The fluid antennas on the FAS should satisfy $0 \leq x_i\leq L$ for $i\in\{1,2\}$ and the minimum spacing requirement. For the former, we have
\begin{equation}\label{bq24}
0 \leq u+v \leq L \text{ and } 0 \leq u-v \leq L
\end{equation}
by using \eqref{bq13}. Because $u>v>0$, \eqref{bq24} is equivalent to 
\begin{equation}\label{bq25}
0< v < u \leq L - v.
\end{equation}
For the minimum spacing requirement, we have
\begin{equation}\label{bq26}
v\geq \frac{\lambda}{4}.
\end{equation}

\subsection{The Condition of $K>3$}

When $K>3$, the interference alignment equations are given by \eqref{bq1}. Assume there exist $M=2^P$ fluid antennas in the FAS. Inspired by \eqref{bq13}, by denoting
\begin{equation}
m=\sum_{p=1}^{P}\frac{b_p+1}{2}\cdot2^{p-1}+1
\end{equation}
where $b_p\in\{-1,1\}$ for $p\in\{1,2,\cdots,P\}$, we express the position coordinate of the $m$th fluid antenna as \cite{LZhu}
\begin{equation}\label{cq2}
x_{m} = u + \sum_{p=1}^P b_p v_p
\end{equation}
where $x_M>x_{M-1}>\cdots>x_1$ and $u>v_P>\cdots>v_1>0$ are assumed. Using \eqref{cq2}, we have the following proposition.

\emph{Proposition 1}: The left-hand side of \eqref{bq1} is
\begin{equation}\label{cq3}
\sum_{m=1}^M \cos\left(\psi_{i} + \alpha_{i} x_m \right)= 2^P \cos(\psi_i + \alpha_i u) \prod_{p=1}^P \cos(\alpha_i v_p) \end{equation}
for $i\in\{1,\cdots,K-1\}$.

\emph{Proof}: See Appendix A. $\hfill\blacksquare$

Using Proposition 1, the interference alignment equations are equivalently transformed into
\begin{equation}
\cos(\psi_i + \alpha_i u) \prod_{p=1}^P \cos(\alpha_i v_p)=0
\end{equation}
for $i\in\{1,\cdots,K-1\}$. Without loss of generality, we consider the following case
\begin{equation}\label{cq5}
\left\{\begin{array}{l}
\cos(\psi_1 + \alpha_1 u) = 0,\\
\cos(\alpha_i v_{i-1}) = 0,\ i = 2, 3, \cdots, K-1.
\end{array}
\right.
\end{equation}
From \eqref{cq5}, to satisfy the interference alignment equations requires
\begin{equation}\label{cq7}
P\geq(K-2), \text{ i.e., } M\geq2^{K-2}.
\end{equation}
The solution to \eqref{cq5} is
\begin{align}\label{cq8}
u=&\frac{n_1 \pi - \psi_1+\pi/2}{\alpha_1},\\
\label{cq9}v_{i-1}= &\frac{n_{i}\pi+\pi/2}{\alpha_{i}},\ i = 2, 3, \cdots, K-1
\end{align}
where $n_1,n_2,\cdots,n_{K-1}\in\mathbb{Z}$ are integers. 

Similar to \eqref{bq25} and \eqref{bq26}, $u$ and $v_p$, $p\in\{1,2,\cdots,K-2\}$, should satisfy
\begin{equation}
0< V < u \leq L - V
\end{equation}
where $V=\sum_{p=1}^P v_p$, which implies 
\begin{equation}
V\leq \frac{L}{2}.
\end{equation}
Furthermore, for any two distinct fluid antennas $m \neq m'$ where $m'=\sum_{p'=1}^{P}\frac{b_{p'}+1}{2}\cdot2^{p'-1}+1$, their distance should satisfy the minimum spacing constraint
\begin{equation}
|x_{m} - x_{m'}| = 2\left| \sum_{p=1}^{P} d_{p} v_{p} \right| \ge \frac{\lambda}{2}
\end{equation}
where $d_p=\frac{1}{2}(b_p-b_{p'})\in\{-1,0,1\}$ for $p\in\{1,2,\cdots,P\}$.

\section{Outage Probability Analysis}

In this letter, the outage event is defined as the failure to satisfy the interference alignment equations \eqref{bq1} given random channel parameters. For the random channels, we assume
\begin{equation}
\psi_{i}, \theta_{i}\sim \mathcal{U}(0,2\pi)
\end{equation}
for $i\in\{1,\cdots,K-1\}$ where $\theta_{i}$ is the abbreviation for $\theta_{i,K}$. To continue, we have the following proportion.

\emph{Proposition 2}: In \eqref{cq8} and \eqref{cq9}, since $n_1,n_2,\cdots,n_{K-1}\in\mathbb{Z}$, by choosing the proper values of $n_1,n_2,\cdots,n_{K-1}$, the outage probability is only related with $|\alpha_{i}|= \frac{2\pi}{\lambda}|\cos\theta_{i}|$.

\emph{Proof}: Since $v_{i}>0$ for $i\in\{1,\cdots,P\}$, from \eqref{cq9}, if $\alpha_{i,k}>0$, we have $n_i\geq0$. If $\alpha_{i,k}<0$, we have $n_i\leq-1$. Thus, the feasibility of $v_{i}>0$ is only related with $|\alpha_{i}|$. Similarly, the feasibility of $u>0$ is only related with $|\alpha_{i}|$. The outage probability is the probability of infeasible value of $v_{i}>0$ to satisfy \eqref{cq9} or infeasible value of $u>0$ to satisfy \eqref{cq8}. Therefore, we have Proposition 2.$\hfill\blacksquare$

Using Proposition 2, since $\theta_{i}\sim \mathcal{U}(0,2\pi)$, we obtain the probability density function (PDF) of $\tau_i=|\cos\theta_{i}|$ as follows
\begin{equation}
f_{\tau_i}(t)=\frac{2}{\pi}\frac{1}{\sqrt{1-t^2}}.
\end{equation}

\subsection{The Condition of $M=2$ and $K=3$}

We first consider Case 1. From \eqref{bq20}, using Proposition 2, we have
\begin{equation}\label{cq20}
v = \frac{n_2\pi+\pi/2}{|\alpha_2|} = \frac{\lambda}{2\tau_2}\left(n_2+\frac{1}{2}\right)
\end{equation}
where $n_2\in\mathbb{Z}_{\geq0}$. Furthermore, from the minimum spacing requirement \eqref{bq26}, the minimum feasible value of $n_2$ is $n_{2}=0$. The corresponding minimum feasible value of $v$ is
\begin{equation}
v_{\min}=\frac{\lambda}{4\tau_2}
\end{equation}
From \eqref{bq25}, if there exists feasible value of $u$, we know that the value of $v$ satisfies $v\le L/2$. Thus, we have $v_{\min}\le L/2$.

In the following, we discuss the feasibility of $u$. From \eqref{bq20}, all possible values of $u$ with $n_1\in\mathbb{Z}$ have the minimal interval
\begin{equation}
\Delta= \frac{\pi}{|\alpha_1|} = \frac{\lambda}{2\tau_1}.
\end{equation}
When $v$ is given, from \eqref{bq25}, $u$ should fall within $(v, L-v]$, whose length is $L-2v$. When $\Delta\leq L-2v$, we can always find at least a feasible $n_1\in\mathbb{Z}$ such that $u$ falls within $(v, L-v]$. When $\Delta>L-2v$, because $\psi_1\sim \mathcal{U}(0,2\pi)$, we have
\begin{equation}
\frac{|\psi_1|}{|\alpha_1|}\sim \mathcal{U}(0,2\Delta)
\end{equation}
and thus the probability to find a feasible $u$ falls within $(v, L-v]$ is $(L-2v)/\Delta$. Combining the aforementioned two situations, the probability to find a feasible $u$ falls within $(v, L-v]$ is $\min\{1,(L-2v)/\Delta\}$. The corresponding outage probability is $\max\{0,1-(L-2v)/\Delta\}$.

Given $\tau_1$ and $\tau_2$, we select $v = v_{\min}$ which maximizes $L-2v$ and thus maximizes the success probability to find a feasible $u$ falls within $(v, L-v]$. Under this condition, $v\le L/2$ is equivalent to $\tau_2\geq\frac{\lambda}{2L}$. Therefore, the outage probability of Case 1 is  
\begin{equation}
p_{1}=\left\{\begin{array}{lc}
\max\left\{0,1 -\frac{2L\tau_1}{\lambda}+\frac{\tau_1}{\tau_2}\right\}; & \text{if } \tau_2\geq\frac{\lambda}{2L}, \\
1; & \text{otherwise}.
\end{array}\right.
\end{equation}

Similarly, the outage probability of Case 2 is
\begin{equation}
p_{2}=\left\{\begin{array}{lc}
\max\left\{0,1 -\frac{2L\tau_2}{\lambda}+\frac{\tau_2}{\tau_1}\right\}; & \text{if } \tau_1\geq\frac{\lambda}{2L}, \\
1; & \text{otherwise}.
\end{array}\right.
\end{equation}

The outage of IA occurs if and only if both Case 1 and Case 2 fail. Given $Z_1$ and $Z_2$, the success or failure of Case 1 depends solely on $\psi_1$ while that of Case 2 depends solely on $\psi_2$. Since $\psi_1$ and $\psi_2$ are independent with each other, the outage probability of IA is
\begin{equation}
P_{\text{out}} = \Bigl(\frac{2}{\pi}\Bigr)^{2}
\int_{0}^{1} \int_{0}^{1}
\frac{p_1 p_2}
{\sqrt{(1-\tau_{1}^{2})\cdot(1-\tau_{2}^{2})}}
d\tau_{1} d\tau_{2}.
\end{equation}

\subsection{The Condition of $K>3$}

In this subsection, we consider the minimal number of fluid antennas equipped at each FAS, i.e., from \eqref{cq7},
\begin{equation}
P=(K-2) \text{ and } M=2^{K-2}.
\end{equation}
The total number of interferers is $K-1$. Assume that we select the $k$th interferer to align via $u$, and assign the remaining interferers, specified by 
\begin{equation}
\bm{\pi}^k=(\pi_1^k, \pi_2^k, \cdots, \pi_{K-2}^k)
\end{equation}
to $(v_{1},v_{2},\cdots,v_{K-2})$, respectively, where 
\begin{equation}
\pi_p^k\in\{1,\cdots,K-1\}\setminus\{k\}
\end{equation}
and $\pi_{p_1}^k\neq \pi_{p_2}^k$ for $p,p_1,p_2\in\{1,\cdots,K-2\}$. Similar to \eqref{cq20}, we have
\begin{equation}
v_p= \frac{\lambda}{2\tau_{\pi_{p}^k}}\left(n_p+\frac{1}{2}\right)
\end{equation}
where $n_p\in\mathbb{Z}_{\geq0}$. By defining
\begin{align}
V^k=\min_{\bm{\pi}^k,n_{1},\cdots, n_{K-2}\in\mathbb{Z}_{\geq0}}&\sum_{p=1}^P v_p \ \
\text{s.t.}\left| \sum_{p=1}^{P} d_{p} v_{p} \right| \ge \frac{\lambda}{4},
\end{align}
the outage probability of this case is  
\begin{equation}
p_{k}=\left\{\begin{array}{lc}
\max\left\{0,1 -\frac{2\tau_k}{\lambda}(L-2V^k)\right\}; & \text{if } V^k\leq \frac{L}{2},\\
1; & \text{otherwise}.
\end{array}\right.
\end{equation}
Therefore, the outage probability of IA is
\begin{equation}
P_{\text{out}}=\left(\frac{2}{\pi}\right)^{K-1}
\int_0^1\cdots\int_0^1\frac{\prod_{k=1}^{K-1}p_k}{\prod_{k=1}^{K-1}\sqrt{1-\tau_{i}^{2}}}d\tau_{1} \cdots d\tau_{K-1}.
\end{equation}

\section{Simulation Results}

In simulations, we assume that each receiver is equipped with an FAS that consists of $M=2^{K-2}$ fluid antennas. The complex propagation gain at the position with coordinate of 0, $g_{i,k}$, $i,k\in\mathcal{K}$, is generated as $\mathcal{CN}(0,1)$. Since $\mathbb{E}[|s_i|^2]=1$, the signal-to-noise ratio (SNR) is defined as $1/\sigma^2$.
 
In Fig. 1, we compare the average achievable sum rates our proposed FAS-enabled interference alignment scheme, the scheme without interference alignment, and the conventional single-user transmission over non-interfering channels, denoted as ``IA", ``w/o IA", and ``Non-Interfering" in the legend, respectively, where $K=3$ and $M=2$. In the scheme without interference alignment, the antennas are uniformly deployed along the FAS. From Fig. 1, it is observed that our proposed interference alignment scheme achieves a substantially better performance than the scheme without IA. Although the proposed scheme is inferior to conventional single-user transmission over non-interfering channels under $L=2\lambda$, the former greatly outperforms the latter when $L=4\lambda$ and $L=6\lambda$.

\begin{figure}
\centering
\includegraphics[width=3.6in]{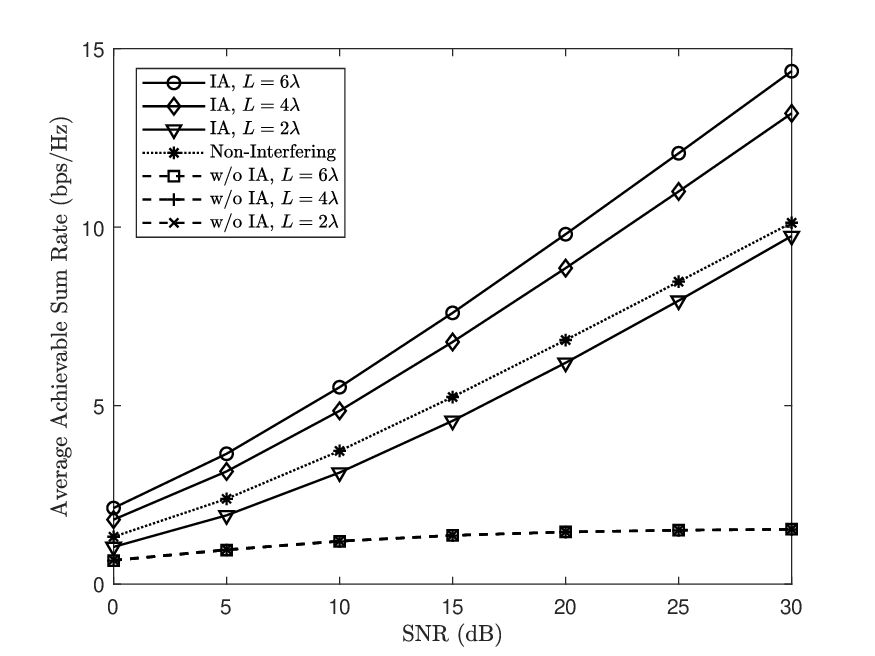}
\caption{Average achievable sum rate versus SNR; comparison of our proposed FAS-enabled interference alignment scheme, the scheme without interference alignment, and the conventional single-user transmission over non-interfering channels, where $K=3$ and $M=2$.}
\end{figure}

In Fig. 2, we present the theoretical and simulated outage probabilities of our proposed FAS-enabled interference alignment scheme, denoted as ``Theoretical" and ``Simulated" in the legend, respectively. From Fig. 2, it is found that with the increase of $L/\lambda$, the outage probabilities decrease. Furthermore, the theoretically derived outage probability matches the simulated results.

\begin{figure}
\centering
\includegraphics[width=3.6in]{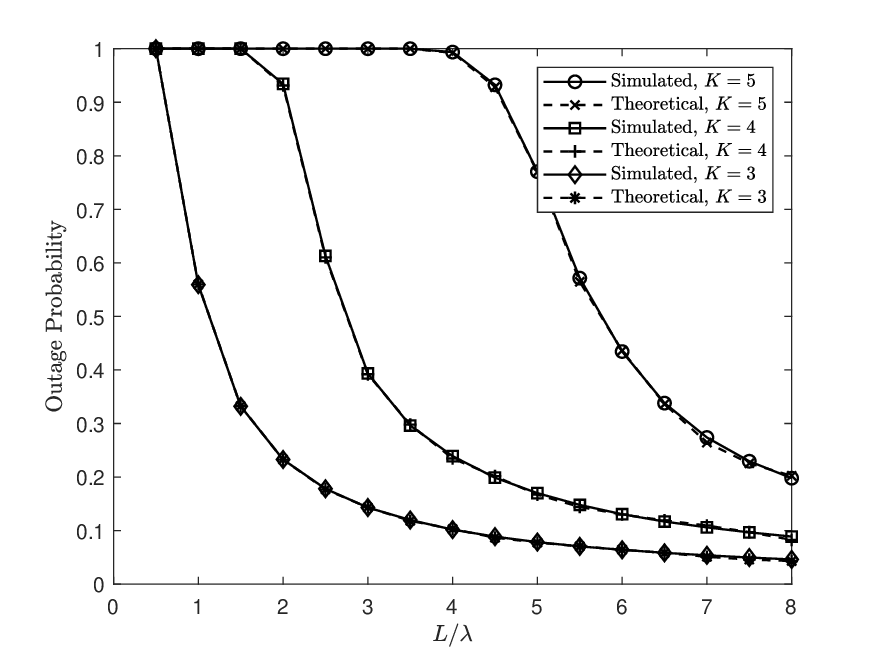}
\caption{Outage Probability versus $L/\lambda$; the theoretical and simulated outage probabilities of our proposed FAS-enabled interference alignment scheme.}
\end{figure}

\section{Conclusion}

In this letter, assuming no channel state information at the transmitters, we have proposed an FAS-enabled interference alignment scheme for a multiuser interference channel. We have also theoretically derived the interference alignment conditions and corresponding outage probabilities. It is illustrated through simulations that the theoretically derived outage probabilities match the numerical results. It is also found that the proposed system can outperform conventional single-user transmission schemes over non-interfering channels.

\appendices
\section{Proof of Proposition 1}

Using \eqref{cq2}, we have
\begin{align}\label{aq1}
\sum_{m=1}^M e^{j(\psi_i + \alpha_i x_{m})}= e^{j(\psi_i + \alpha_i u)} \sum_{b_1,b_2,\cdots,b_P} \prod_{p=1}^P e^{j\alpha_i b_p v_p}.
\end{align}
Since $b_p\in\{-1,1\}$ for $p\in\{1,2,\cdots,P\}$, we obtain
\begin{align}\label{aq2}
\sum_{b_1,b_2,\cdots,b_P} \prod_{p=1}^P e^{j\alpha_i b_p v_p}=&\prod_{p=1}^P \left( e^{j\alpha_i v_p} + e^{-j\alpha_i v_p} \right)\nonumber\\
&=\prod_{p=1}^P 2 \cos(\alpha_i v_p).
\end{align}
Substituting \eqref{aq2} into \eqref{aq1} and taking its real part yields \eqref{cq3}.

\end{document}